\documentclass[aps,pre,twocolumn,showpacs,preprintnumbers,amsmath,amssymb]{revtex4}
\usepackage{graphicx}
\usepackage{dcolumn}
\usepackage{bm}
\usepackage{color}
\usepackage{texdraw}
\begin{document}
\title{Modes of nonlinear acoustic transparency in the strained 
paramagnetic crystal}
\author{S.V. Sazonov}
 \email{barab@newmail.ru}
 \affiliation{I. Kant State University of Russia, 236041 Kaliningrad, Russia}
\author{N.V. Ustinov}
 \email{n_ustinov@mail.ru}
 \affiliation{Tomsk State University, 634050 Tomsk, Russia }
\date{\today}
\begin{abstract}
The propagation of the transverse-longitudinal acoustic pulses through a 
strained cubic crystal containing the resonant paramagnetic impurities with 
effective spin $S=1$ is investigated. 
It is supposed that the pulses propagate under arbitrary angle with respect to 
the direction of the external static deformation parallel to the fourth-order 
symmetry axis. 
In this geometry, both the transverse and longitudinal components of the 
acoustic field have the high-frequency and zero-frequency spectral components. 
We show that a pulse can propagate in the modes different from the acoustic 
self-induced transparency. 
In particular, a pulse propagating in the mode of acoustic self-induced 
supertransparency substantially changes the populations of the spin sublevels, 
but its group velocity remains almost equal to the linear velocity of the 
sound. 
If a pulse propagates in the acoustic extraordinary transparency mode, then 
its group velocity is substantially lower while the sublevel populations 
remain virtually invariant. 
Also, the modes of propagation under conditions of weakly excited spin 
transitions and large detuning of the pulse high-frequency components are 
identified. 
\end{abstract}
\pacs{43.25.+y, 43.35.+d}
\maketitle

\section{INTRODUCTION}

Historically, the coherent optical effects found their acoustic analogues 
after a while.
Such a correspondence is easily traced, for instance, in the case of the 
self-induced transparency (SIT) phenomenon. 
Its discovery in 1967 \cite{McH1} and theoretical explanation in 1969 
\cite{McH2} stimulated a search for the acoustic analogue. 
The experimental observation at the liquid helium temperature of the acoustic 
self-induced transparency (ASIT) of the longitudinal hypersound in cubic 
crystal $\rm MgO$ containing paramagnetic ions $\rm Fe^{2+}$ and $\rm Ni^{2+}$ 
was carried out in 1970 by Shiren \cite{Sh}, who gave also the theoretical 
consideration of this effect. 
Independently, ASIT was studied theoretically for transverse hypersound in a 
system of paramagnetic impurities with effective spin $S=1/2$ \cite{D}. 
Soon after that, this effect was successfully revealed in a crystal 
$\rm LiNbO_3$ alloyed by ions $\rm Fe^{2+}$ \cite{SSSh}. 

Despite the existence of the optical-acoustic analogies mentioned above, the 
acoustic coherent effects do not appear as the exact copies of optical ones. 
Obviously, the quantitative differences connected with a great distinction 
of the light and sound velocities or the frequencies of the electric-dipole 
and spin-phonon transitions take place. 
It should be stressed that the qualitative differences exist there. 
Thus, the optical wave is especially transverse whereas the acoustic one has 
the longitudinal-transverse structure. 

The interaction between transverse and longitudinal components of the acoustic 
pulse in a crystal without anharmonicity of the lattice modes happens due to a 
presence of paramagnetic impurities. 
Its influence on the dynamics of the components is inessential, if the linear 
velocities of acoustic waves differ significantly, as it takes place in the 
most solids. 
However, the linear velocities of transverse and longitudinal sounds are close 
in the ion crystals of halogenide of alkaline metals \cite{TR}. 
Since the components of the strain field interact effectively in this case, 
new features of their dynamics have to be expected. 
Indeed, new mode of the resonant acoustic transparency has been revealed 
theoretically in Ref.~\onlinecite{VS}. 
The propagation of the transverse-longitudinal elastic pulses in this mode 
is accompanied by trapping of the population of the spin sublevels. 
Nevertheless, the deceleration in the velocity of the pulses is comparable to 
that in the case of ASIT of especially longitudinal or transverse 
hypersound \cite{Sh,D}.

The theoretical considerations of the nonlinear dynamics of  
transverse-longitudinal acoustic pulses were fulfilled basically for the 
paramagnetic ions with effective spin $S=1/2$ \cite{VS,Z1,Z2}. 
At the same time, it is well-known that the strongest interaction with the 
crystal lattice oscillations takes place in the case $S=1$ \cite{TR}. 
Since the spin-phonon interaction is greater here on two or three orders than 
for $S=1/2$, the investigation of acoustic transparency of 
transverse-longitudinal pulses in a system of paramagnetic impurities with 
effective spin $S=1$ is more preferable from the point of view of experimental 
testing. 

The resonant modes of propagation of the transverse-longitudinal hypersound 
in direction parallel to external magnetic field and an axis of symmetry of 
the fourth order of the paramagnetic crystal with $S=1$ have been studied in 
\cite{GS}. 
It was shown that these modes coincide with ones discussed in 
Ref.~\onlinecite{VS}. 
The distinctive feature of the geometry considered there is that the roles of 
transverse and longitudinal components of the acoustic pulses are strictly 
various: the former component excites the spin-phonon transitions in the 
Zeeman triplet, the latter one causes the dynamic shifts of the frequencies of 
these transitions. 

Recently, it has been shown in the case of especially longitudinal acoustic 
pulses propagating under arbitrary angle with respect to an axis of the 
symmetry of the paramagnetic crystal, that both functions can be carried out 
by a single component of the strain field \cite{SU1}. 
In this connection, the studying of the nonlinear modes of the acoustic 
transparency for the transverse-longitudinal pulses that propagate in a system 
of the effective spins $S=1$ under arbitrary direction to the symmetry axis is 
of great interest. 
The present report is devoted to allocating these modes. 
Following to Ref.~\onlinecite{SU1}, we suppose here that the external magnetic 
field is absent, and splitting of the energy sublevels is created by the 
static deformation of the paramagnetic crystal. 
For this reason, the degeneration of the spin states is removed incompletely 
in contrast to usual ASIT. 

The paper is organized as follows.
In Section II, the system of material and wave equations describing the 
interaction in the strained crystal of the transverse-longitudinal elastic 
pulses with the paramagnetic impurities having effective spin $S=1$ is 
derived. 
We use the semi-classical approach, i.e., the dynamics of the acoustic fields 
is assumed to obey the classical Hamilton equations for the continuous medium, 
but the spin subsystem is treated as a quantum object. 
The order of the derivatives in the wave equations on the strain field 
components is reduced with the help of the unidirectional propagation (UP) 
approximation. 
In Section III, the approximation of slowly varying envelopes (SVE) is applied 
to the particular case of the system obtained in the previous section. 
It is shown that the evolution of the acoustic pulses in the paramagnetic 
crystal is governed by the well-studied equations of the long/short-wave 
coupling (LSWC) \cite{S1,SU2}. 
The pulse solutions decreasing exponentially and rationally of the LSWC 
equations are presented in Section IV. 
These solutions are exploited to classify the modes of the acoustic 
transparency of transverse-longitudinal pulses in the deformed paramagnetic 
crystal. 

\section{BASIC MODEL}

Consider a cubic crystal containing the paramagnetic impurities with effective 
spin $S=1$. 
Let a transverse-longitudinal acoustic pulse propagate in the crystal along 
the $z$ axis directed under arbitrary angle $\alpha$ with respect to one of 
its axes of symmetry of the fourth-order ($z'$ axis; see Fig.~1). 
\begin{figure}[ht]
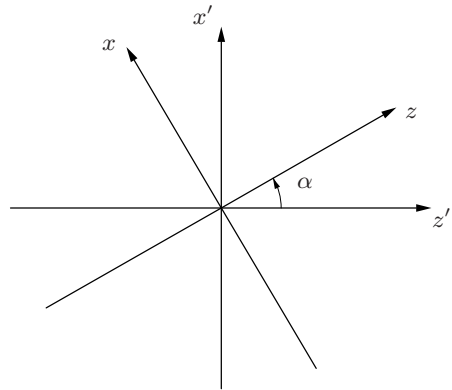

\vskip0.3cm
\centertexdraw{
\drawdim cm
\linewd 0.02
\arrowheadtype t:F \arrowheadsize l:0.2 w:0.1 
\move(0 2.4) \avec(5.6 2.4) \textref h:L v:T \htext{$z'$}
\move(2.8 -0.01) \avec(2.8 4.81) \move(2.8 4.85) \textref h:R v:B \htext{$x'$\ }
\move (2.8 2.4) \avec(5.13 3.73) \textref h:L v:T \htext{\ $z$}
\move (2.8 2.4) \lvec(0.47 1.07)
\move (2.8 2.4) \avec(1.54 4.54) \move(1.4 4.5) \textref h:R v:B \htext{$x$}
\move (2.8 2.4) \lvec(4.06 0.26)
\move (2.8 2.4) \linewd 0.01
\larc r:0.8 sd:0 ed:30.0
\move (3.55 2.67) \setgray 0 \arrowheadtype t:F \arrowheadsize l:0.14 w:0.08 
\avec(3.49 2.79)
\move (3.8 2.8) \textref h:L v:T \htext{$\alpha$}
\setgray 1 \move(-0.4 -0.1) \lvec(-0.4 5) \move(6.0 -0.1) \lvec(6.0 5)
}
\caption{Scheme of the coordinate axes.}
\end{figure}
Also, we assume that an external field of static and uniform deformation 
employing the crystal is parallel to the $z'$ axis. 

The Hamiltonian $\hat H$ of the spin-elastic interaction that arises in the 
most general case \cite{S2} is a function of the bilinear combination of spin 
operators $\hat S_j$ ($j=x',y,z'$): 
\begin{eqnarray}
\hat H=f(\hat S_iQ_{ij}\hat S_j). 
\label{H_g}
\end{eqnarray}
Coefficients $Q_{ij}$ in this formula are supposed to depend on the components 
of the strain tensor 
\[
{\cal E}_{ml}=\frac12\left(\frac{\partial U_m}{\partial x_l}
+\frac{\partial U_l}{\partial x_m}\right) 
\]
($m,l=x',y,z'$; $U_m$ are the components of the displacement vector $\bf U$). 
In a representation of the eigenfunctions of operator $\hat S_{z'}$, the spin 
matrices take the form \cite{TR} 
\begin{eqnarray}
\begin{array}{c}
{\hat S}_{x'}=\frac{1}{\sqrt2}\left(
\begin{array}{ccc}
0&1&0\\1&0&1\\0&1&0
\end{array}
\right),\quad
{\hat S}_{y}=\frac{i}{\sqrt2}\left(
\begin{array}{ccc}
0&-1&0\\1&0&-1\\0&1&0
\end{array}
\right),\\
{\hat S}_{z'}=\left(
\begin{array}{ccc}
1&0&0\\0&0&0\\0&0&-1
\end{array}
\right).
\end{array}
\label{S_j}
\end{eqnarray}

The coefficients of the bilinear combination satisfy condition 
$Q_{ij}=\delta_{ij}$ ($\delta_{ij}$ is the Kronecker symbol) if the external 
deformation of the crystal is absent. 
Expanding these coefficients and function $f$ in the power series on 
${\cal E}_{ml}$ and retaining the first order terms, we obtain 
\[
f=f({\hat S}^2)+f^{\prime}({\hat S}^2)\left(\frac{\partial Q_{ij}}
{\partial {\cal E}_{ml}}\right)_0\hat S_i\hat S_j{\cal E}_{ml}, 
\]
where subscript ''0'' means that the values of derivatives correspond to 
nondeformed crystal. 
The first term depending on the Kazimir operator ${\hat S}^2=S(S+1)=2$ is 
omitted in what follows as the constant addition to $\hat H$. 
Then, Hamiltonian (\ref{H_g}) is expressed through the components of the 
spin-elastic interaction tensor 
\[
G_{ijlm}=f^{\prime}({\hat S}^2)\left(\frac{\partial Q_{ij}}
{\partial {\cal E}_{ml}}\right)_0 
\]
as given 
\[
\hat H=\frac12G_{ijlm}{\cal E}_{ml}(\hat S_i\hat S_j+\hat S_j\hat S_i). 
\]

In the case of the geometry we consider, 
the last equation is rewritten in the following manner 
\[
\hat H=\hat H_0+\hat V,  
\]
where $\hat H_0$ is the Hamiltonian of the effective spin in the field 
${\cal E}^{(0)}_{z'z'}$ of the external deformation: 
\begin{eqnarray}
\hat H_0=G_{||}{\cal E}^{(0)}_{z'z'}{\hat S}^2_{z'}=\hbar\omega_0
\left(
\begin{array}{ccc}
1&0&0\\0&0&0\\0&0&1
\end{array}
\right),
\label{H_0}
\end{eqnarray}
$\hat V$ is the Hamiltonian of interaction between the effective spin and 
strain field of the acoustic pulse propagating through the crystal \cite{S2}: 
\begin{eqnarray}
\hat V=G_{||}{\cal E}_{zz}{\hat S}^2_{z}+\frac{G_{\perp}}{2}\left[
{\cal E}_{xz}({\hat S}_{x}{\hat S}_{z}+{\hat S}_{z}{\hat S}_{x})
\right.\nonumber\\ \left.\mbox{}
+{\cal E}_{yz}({\hat S}_{y}{\hat S}_{z}+{\hat S}_{z}{\hat S}_{y})\right]\!.
\label{V}
\end{eqnarray}
Here we use notations $G_{||}=G_{z'z'z'z'}=G_{x'x'x'x'}=G_{yyyy}$,
$\omega_0=G_{||}{\cal E}^{(0)}_{z'z'}/\hbar$ ($\hbar$ is the Planck constant) 
and $G_{\perp}=G_{x'x'z'z'}=G_{x'x'yy}=G_{yyz'z'}$. 
Operators ${\hat S}_{x}$ and ${\hat S}_{z}$ are connected with ${\hat S}_{x'}$ 
and ${\hat S}_{z'}$ by means of the transformation of rotation \cite{Sh,JS}
\begin{eqnarray}
\begin{array}{c}
{\hat S}_{x_{\mathstrut}}={\hat S}_{x'}\cos\alpha-{\hat S}_{z'}\sin\alpha,\\
{{\hat S}_z}^{\mathstrut}={\hat S}_{z'}\cos\alpha+{\hat S}_{x'}\sin\alpha.
\end{array}
\label{rot}
\end{eqnarray}

For the consideration of the dynamics of the effective spins and the acoustic 
pulse to be self-consistent, we introduce the Hamiltonian of the elastic field 
\begin{eqnarray}
H_a=\frac12\int\left\{\frac{p_x^2+p_y^2+p_z^2}{\rho}+
\rho a_{||}^2\left(\frac{\partial U_z}{\partial z}\right)^2
\right.\nonumber\\ \left.\mbox{}
+\rho a_{\perp}^2\left[\left(\frac{\partial U_x}{\partial z}\right)^2+
\left(\frac{\partial U_y}{\partial z}\right)^2\right]\right\}d{\bf r}, 
\label{H_a}
\end{eqnarray}
where $\rho$ is the mean crystal density; $p_x$, $p_y$ and $p_z$ are the 
components of momentum density of the local displacement of the crystal; 
$a_{||}$ and $a_{\perp}$ are linear velocities of the longitudinal and 
transversal hypersounds, respectively. 
We assume in (\ref{H_a}) that all dynamical displacements $U_x$, $U_y$ and 
$U_z$ depend on variables $z$ and $t$ only. 
The integral is taken over the crystal volume. 

In accordance with the general scheme of the semi-classical approach 
\cite{S2,S3}, we describe evolution of the effective spin by the equation on 
the density matrix $\hat\rho$\,: 
\begin{eqnarray}
i\hbar\frac{\partial\hat\rho}{\partial t}=[\hat H_0+\hat V,\hat\rho].
\label{rho_t}
\end{eqnarray}
At the same time, dynamics of the acoustic field obey the classical Hamilton 
equations for the continuous medium: 
\begin{eqnarray}
\begin{array}{c}
\displaystyle\frac{\partial\bf p}{\partial t}=-
{\frac{\delta}{\delta\bf U}}_{\mathstrut}
\left(H_a+<\!\hat {\tilde V}\!>\right),\\
\displaystyle\frac{\partial\bf U}{\partial t}=
{\frac{\delta}{\delta\bf p}}^{\mathstrut}
\left(H_a+<\!\hat {\tilde V}\!>\right),
\end{array}
\label{Ham}
\end{eqnarray}
where
\[
\displaystyle<\!\hat {\tilde V}\!>\,=\!\int\! n\!<\!\hat V\!>\!d{\bf r}, 
\]
$n$ is the concentration of the paramagnetic impurities;  
$<\!\hat V\!>\,={\rm Tr}\,(\hat\rho\hat V)$ is the quantum average of 
$\hat V$. 

Let the density matrix be represented in the next form 
\begin{eqnarray}
\hat\rho=\left(
\begin{array}{ccc}
\rho_{33}&\rho_{32}&\rho_{31}\\
\rho_{23}&\rho_{22}&\rho_{21}\\
\rho_{13}&\rho_{12}&\rho_{11}
\end{array}
\right). 
\label{rho}
\end{eqnarray}
Eqs. (\ref{S_j}), (\ref{V}) and (\ref{rot}) give us the following expression 
for the Hamiltonian of the spin-phonon interaction: 
\begin{eqnarray}
\hat V=\left(
\begin{array}{ccc}
V_{11}&-V^*_{12}&V^*_{13}\\
-V_{12}&V_{22}&V^*_{12}\\
V_{13}&V_{12}&V_{11}
\end{array}
\right),
\label{H}
\end{eqnarray}
where 
\[
V_{11}=\frac{G_{||}}{2}(1+\cos^2\!\alpha)\,{\cal E}_{zz}-
\frac{G_{\perp}}{2}\sin2\alpha\,\,{\cal E}_{xz},
\]
\[
V_{22}=G_{||}\sin^2\!\alpha\,\,{\cal E}_{zz}+
G_{\perp}\sin2\alpha\,\,{\cal E}_{xz},
\]
\[
V_{12}=-\frac{G_{||}\sin2\alpha}{2\sqrt2}\,\,{\cal E}_{zz}
-\frac{G_{\perp}}
{\sqrt2}(\cos2\alpha\,\,{\cal E}_{xz}+i\cos\alpha\,\,{\cal E}_{yz}),
\]
\[
V_{13}=\frac{G_{||}}{2}\sin^2\!\alpha\,\,{\cal E}_{zz}
+G_{\perp}(\sin2\alpha\,\,{\cal E}_{xz}/2+i\sin\alpha\,\,{\cal E}_{yz}). 
\]

The physical mechanism of spin-elastic interaction in the case considered is 
the Van Vleck mechanism \cite{TR}. 
The strain fields modulate the intracrystalline electric field in a location 
of the paramagnetic ions. 
Due to the quadrupole Stark effect, the static gradient of the electric field 
causes a splitting of the quantum sublevels of effective spin $S=1$ that occur 
to be degenerate on the absolute value of its projection $S_{z'}$ (see 
(\ref{H_0}) and Fig.~2). 
\begin{figure}[ht]
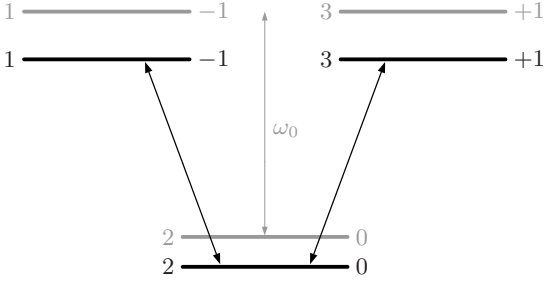

\vskip-0.7cm
\centertexdraw{
\drawdim cm
\linewd 0.05
\move(2.9 0.6) \textref h:R v:C \htext{$2$\ }
\lvec(5.1 0.6) \textref h:L v:C \htext{\ $0$}
\move(0.8 3.359) \textref h:R v:C \htext{$1$\ }
\lvec(3 3.359) \textref h:L v:C \htext{\ $-1$}
\move(5 3.359) \textref h:R v:C \htext{$3$\ }
\lvec(7.2 3.359) \textref h:L v:C \htext{\ $+1$}
\setgray 0.6
\move(2.9 1) \textref h:R v:C \htext{\textcolor[rgb]{0.6,0.6,0.6}{$2$}\ }
\lvec(5.1 1) \textref h:L v:C \htext{\ \textcolor[rgb]{0.6,0.6,0.6}{$0$}}
\move(0.8 4.0) \textref h:R v:C \htext{\textcolor[rgb]{0.6,0.6,0.6}{$1$}\ }
\lvec(3 4.0) \textref h:L v:C \htext{\ \textcolor[rgb]{0.6,0.6,0.6}{$-1$}}
\move(5 4.0) \textref h:R v:C \htext{\textcolor[rgb]{0.6,0.6,0.6}{$3$}\ }
\lvec(7.2 4.0) \textref h:L v:C \htext{\ \textcolor[rgb]{0.6,0.6,0.6}{$+1$}}
\linewd 0.008
\move(4 2) \arrowheadtype t:F \arrowheadsize l:0.12 w:0.08 \avec(4 4)
\move(4 2) \arrowheadtype t:F \arrowheadsize l:0.12 w:0.08 \avec(4 1.02)
\move(4 2.5) \textref h:L v:T \htext{\textcolor[rgb]{0.6,0.6,0.6}{\ $\omega_0$}}
\setgray 0
\linewd 0.02
\move(2.9 1.97885)
\arrowheadtype t:F \arrowheadsize l:0.14 w:0.1 \ravec(0.5 -1.35)
\move(2.9 1.97885)
\arrowheadtype t:F \arrowheadsize l:0.14 w:0.1 \ravec(-0.5 1.35)
\move(5.1 1.97885)
\arrowheadtype t:F \arrowheadsize l:0.14 w:0.1 \ravec(-0.5 -1.35)
\move(5.1 1.97885)
\arrowheadtype t:F \arrowheadsize l:0.14 w:0.1 \ravec(0.5 1.35)
\setgray 1
\move(0 0) \lvec(0 5) \move(8 0) \lvec(8 5)
}
\vskip-0.4cm
\caption{Splitting of spin sublevels in the field of the static deformation 
(gray color), their position in a presence of the acoustic pulse and quantum 
transitions excited.
The number of the level and the corresponding value of projection $S_{z'}$ of 
effective spin on the axis of the external deformation are indicated from the 
left and right, respectively.}
\end{figure}
The components of the acoustic pulse excite the electro-quadrupole transitions 
between these sublevels and, as it follows from the expressions for 
$V_{11}$ and $V_{22}$, shift dynamically the transition frequency. 

Using (\ref{H_a}) and (\ref{Ham})--(\ref{H}), we deduce the wave equations on 
the strain field components: 
\begin{eqnarray}
\frac{\partial^2{\cal E}_{zz}}{\partial t^2}-a_{||}^2
\frac{\partial^2{\cal E}_{zz}}{\partial z^2}=\frac{nG_{||}}{2\rho}
\frac{\partial^2}{\partial z^2}\Bigl[(3\sin^2\!\alpha-2)\rho_{22}
\nonumber\\ \mbox{}
+\frac{\sin2\alpha}{\sqrt2}(\rho_{23}+\rho_{32}-\rho_{12}-\rho_{21})
\nonumber\\ \mbox{}
+\sin^2\!\alpha(\rho_{13}+\rho_{31})\Bigr],
\label{E_z_tt}
\end{eqnarray}
\begin{eqnarray}
\frac{\partial^2{\cal E}_{xz}}{\partial t^2}-
a_{\perp}^2\frac{\partial^2{\cal E}_{xz}}{\partial z^2}=
\frac{nG_{\perp}}{4\rho}\frac{\partial^2}{\partial z^2}\Bigl[
3\sin2\alpha\,\,\rho_{22}
\nonumber\\ \mbox{}
+\sqrt2\cos2\alpha(\rho_{23}+\rho_{32}-\rho_{12}-\rho_{21})\vphantom{\frac12}
\nonumber\\ \mbox{}
+\sin2\alpha\,(\rho_{13}+\rho_{31})\Bigr],
\label{E_x_tt}
\end{eqnarray}
\begin{eqnarray}
\frac{\partial^2{\cal E}_{yz}}{\partial t^2}-
a_{\perp}^2\frac{\partial^2{\cal E}_{yz}}{\partial z^2}=
i\frac{nG_{\perp}}{2\rho}\frac{\partial^2}{\partial z^2}\Bigl[
\sin\alpha\,(\rho_{31}-\rho_{13})
\nonumber\\ \mbox{}
+\frac{\cos\alpha}{\sqrt2}(\rho_{12}+\rho_{32}-\rho_{21}-\rho_{23})\Bigr].
\label{E_y_tt}
\end{eqnarray}
Equations (\ref{rho_t}) and (\ref{E_z_tt})--(\ref{E_y_tt}) form the 
self-consistent system describing the nonlinear dynamics of 
transverse-longitudinal acoustic pulses in the deformed crystal containing the 
paramagnetic impurities with effective spin $S=1$. 
Since its analysis is very complicated in the general case, we restrict 
subsequent consideration by a specific model. 

Let us assume that the linear velocities of both acoustic waves are equal: 
\[
a_{||}=a_{\perp}=a.
\]
This restriction is rather artificial, since $a_{||}>a_{\perp}$ in solids. 
As noted in the previous section, it is fulfilled best of all in the ion 
crystals of halogenide of alkaline metals \cite{TR}. 
One of the representatives of such the crystals is, for instance, $\rm NaBr$ 
\cite{K}. 
Using typical experimental parameters of medium and acoustic pulses, we show 
at the end of Section IV that strict observance of this condition is not very 
important from physical point of view. 
The case of the acoustic pulse propagation through deformed paramagnetic 
crystal with effective spin $S=1$, in which the linear velocities of the 
components differ essentially, has been studied in Ref.~\onlinecite{SU1}. 
 
As far as the linear velocities coincide, we are able to simplify 
Eqs. (\ref{E_z_tt})--(\ref{E_y_tt}) by reducing them in the order of 
derivative with the help of the UP approximation \cite{EGCB}. 
Indeed, suppose that the concentration of the paramagnetic impurities in the 
crystal is small: 
\[
\eta=\frac{n(G_{\perp}^2+G_{||}^2)}{\hbar\omega_0\rho a^2}\ll1, 
\] 
and introduce new independent variables
\[
\tau=t-\frac{z}{a},\quad\zeta=\eta z, 
\]
which are usually referred to as the retarded time and slow coordinate, 
respectively. 
Obviously, we have
\[
\frac{\partial}{\partial t}=\frac{\partial}{\partial\tau},\qquad
\frac{\partial}{\partial z}=-\frac{1}{a}\frac{\partial}{\partial\tau}+
\eta\frac{\partial}{\partial\zeta}.
\]
Then, neglecting the terms proportional to $\eta^2$, we approximately write 
\[
\frac{\partial^2}{\partial z^2}\approx\frac{1}{a^2}\frac{\partial^2}{\partial
\tau^2}-2\frac{\eta}{a}\frac{\partial^2}{\partial\tau\partial\zeta}
\]
in the left-hand side of Eqs. (\ref{E_z_tt})--(\ref{E_y_tt}) and 
\[
\frac{\partial^2}{\partial z^2}\approx\frac{1}{a^2}\frac{\partial^2}{\partial
\tau^2}
\]
in the right hand side. 
Integration on $\tau$ of the wave equations obtained by this way in the terms 
of new variables gives after returning to the initial ones: 
\begin{widetext}
\begin{eqnarray}
\frac{\partial{\cal E}_{zz}}{\partial z}+\frac{1}{a}
\frac{\partial{\cal E}_{zz}}{\partial t}=\frac{nG_{||}}{4\sqrt2\hbar\rho a^3}
\Bigl[i(\rho_{12}-\rho_{21}+\rho_{23}-\rho_{32})
(\hbar\omega_0\sin2\alpha-2G_{\perp}{\cal E}_{xz})
+2G_{\perp}(\sqrt2\sin\alpha\,(\rho_{33}-\rho_{11})
\nonumber\\ \mbox{}
-\cos\alpha\,(\rho_{12}+\rho_{21}+\rho_{23}+\rho_{32})){\cal E}_{yz}\Bigr],
\label{E_z_z}
\end{eqnarray}
\begin{eqnarray}
\frac{\partial{\cal E}_{xz}}{\partial z}+\frac{1}{a}
\frac{\partial{\cal E}_{xz}}{\partial t}=\frac{nG_{\perp}}{4\sqrt2\hbar\rho a^3}
\Bigl[i(\rho_{12}-\rho_{21}+\rho_{23}-\rho_{32})
(\hbar\omega_0\cos2\alpha+G_{||}{\cal E}_{zz})
+G_{\perp}(\sqrt2\cos\alpha\,(\rho_{33}-\rho_{11})
\nonumber\\ \mbox{}
+\sin\alpha\,(\rho_{12}+\rho_{21}+\rho_{23}+\rho_{32})){\cal E}_{yz}\Bigr],
\label{E_x_z}
\end{eqnarray}
\begin{eqnarray}
\frac{\partial{\cal E}_{yz}}{\partial z}+\frac{1}{a}
\frac{\partial{\cal E}_{yz}}{\partial t}=\frac{nG_{\perp}}
{4\sqrt2\hbar\rho a^3}\Bigl[\sqrt2(\rho_{11}-\rho_{33})
(G_{\perp}\cos\alpha\,{\cal E}_{xz}+G_{||}\sin\alpha\,{\cal E}_{zz})
+(\rho_{12}+\rho_{21}+\rho_{23}+\rho_{32})
\nonumber\\ \mbox{}
\times(\hbar\omega_0\cos\alpha-
G_{\perp}\sin\alpha\,{\cal E}_{xz}+G_{||}\cos\alpha\,{\cal E}_{zz})\Bigr].
\label{E_y_z}
\end{eqnarray}
\end{widetext}
It is assumed hereinafter that the components of the strain field and the 
density matrix elements satisfy following conditions at $t\to-\infty$: 
\[
{\cal E}_{xz,\,yz,\,zz}\to0,\quad\frac{\partial{\cal E}_{xz,\,yz,\,zz}}
{\partial t}\to0,\quad \rho_{jj}\to W_j,\quad\rho_{jk}\to0 
\]
($j,\,k=1,\,2,\,3$; $k\ne j$).
These conditions correspond to the pulsed mode of propagation of the sound in 
a crystal. 
Note that the evolution equation (\ref{rho_t}) on density matrix is used to 
exclude the time derivatives of its elements in the right-hand side of 
Eqs. (\ref{E_z_z})--(\ref{E_y_z}). 

The equations we came to are yet complicated for the rigorous analysis. 
Fortunately, they can be reduced in particular case to a system describing the 
interaction of the transverse-longitudinal acoustic pulses with two-level 
quantum particles. 
The investigation of this system will be carried out in the next sections. 

\section{SYSTEM OF THE LSWC EQUATIONS}

Let the populations of the quantum states with $S_{z'}=\pm1$ be equal: 
\begin{eqnarray}
\rho_{33}=\rho_{11}. 
\label{red_1}
\end{eqnarray}
This assumption is compatible with Eqs. (\ref{rho_t}), 
(\ref{E_z_z})--(\ref{E_y_z}) if following conditions are imposed 
\begin{eqnarray}
\rho_{31}=\rho_{13},\quad\rho_{32}=-{\rho_{12}}_{\mathstrut},\quad
{\cal E}_{yz}={0^{\mathstrut}}^{\mathstrut}. 
\label{red_2}
\end{eqnarray}
It is checked immediately that conditions (\ref{red_2}) are sufficient to keep 
(\ref{red_1}). 
Also, they give us relation 
\begin{eqnarray}
\rho_{13}=W_1-\rho_{11}. 
\label{red_3}
\end{eqnarray}
Taking into account Eqs. (\ref{red_1})--(\ref{red_3}), we rewrite 
(\ref{rho_t}) and (\ref{E_z_z})--(\ref{E_y_z}) in the next form: 
\begin{eqnarray}
\frac{\partial W}{\partial t}=iE_{\perp}(\sigma-\sigma^*), 
\label{s_3_t}
\end{eqnarray}
\begin{eqnarray}
\frac{\partial\sigma}{\partial t}=i(\omega_0+2E_{||})\sigma+2iE_{\perp}W, 
\label{s_t}
\end{eqnarray}
\begin{eqnarray}
\frac{\partial E_{\perp}}{\partial z}+\frac{1}{a}\frac{\partial E_{\perp}}
{\partial t}=-i\frac{nA}{4\hbar\rho a^3}(\omega_0+2E_{||}+\delta E_{\perp})
\nonumber\\ \mbox{}
\times(\sigma-\sigma^*), 
\label{E_t_z}
\end{eqnarray}
\begin{eqnarray}
\frac{\partial E_{||}}{\partial z}+\frac{1}{a}\frac{\partial E_{||}}
{\partial t}=i\frac{nA}{8\hbar\rho a^3}\left[\delta(\omega_0+2E_{||}+\delta 
E_{\perp})
\right.\nonumber\\ \left.\mbox{}
+d^2E_{\perp}\right](\sigma-\sigma^*). 
\label{E_l_z}
\end{eqnarray}
Here
\begin{eqnarray}
\begin{array}{c}
\displaystyle W={\frac{1+W_2}{4}}_{\mathstrut}-\rho_{22},
\quad\sigma=\sqrt2\,\rho_{21},\\
\displaystyle E_{\perp}={\frac{G_{\perp}}{\hbar}}^{\mathstrut}
\cos2\alpha\,\,{\cal E}_{xz}+{\frac{G_{||}}{2\hbar}}_{\mathstrut}\sin2\alpha\,\,
{{\cal E}_{zz}},\\
\displaystyle E_{||}=\frac{G_{||}}{2\hbar}\cos2\alpha\,\,{\cal E}_{zz}-
{\frac{G_{\perp\vphantom{|}}}{\hbar}}^{\mathstrut}\sin2\alpha\,\,{{\cal E}_{xz}}
\end{array}
\label{new_var}
\end{eqnarray}
and
\[
A=G_{\perp}^2\cos^2\!2\alpha+\frac{G_{||}^2}{2}\sin^2\!2\alpha, 
\]
\[
\delta=\frac{2G_{\perp}^2-G_{||}^2}{2A}\sin4\alpha,\quad
d=\frac{\sqrt2}{A}G_{\perp}G_{||}.
\]

System (\ref{s_3_t})--(\ref{E_l_z}) is remarkable from the point of view of 
its physical applications. 
In the case $d=\delta=0$ and $E_{||}=0$, it coincides with well-known reduced 
Maxwell--Bloch equations (RMB) for isotropic two-level medium \cite{EGCB}. 
If $d=0$, then we can put $E_{||}=-\delta E_{\perp}/2$, and the system 
appearing is nothing but the RMB equations for anisotropic medium 
\cite{AEGM}, which describe also the propagation of one-component acoustic 
pulses through the deformed paramagnetic crystals \cite{SU1}. 
At last, Eqs. (\ref{s_3_t})--(\ref{E_l_z}) are equivalent to the system of 
material and wave equations derived in Ref.~\onlinecite{Z1} under 
consideration of the dynamics of transverse-longitudinal acoustic pulses in 
paramagnetic with effective spin $S=1/2$ in the external magnetic field 
presence. 
The complete investigation of this system turned out to be awkward. 
The simplest stationary solution found corresponds to the case $\delta=0$ 
only. 
It describes the propagation of the extremely short acoustic pulse that 
possesses no well-defined carrier frequency in direction parallel to magnetic 
field. 
In contrast to this, the pulses containing of the higher-frequency components 
will be considered below. 

Owing to the last remark, further simplification of system 
(\ref{s_3_t})--(\ref{E_l_z}) will be achieved by applying the SVE 
approximation. 
Comparing the right-hand side of Eqs. (\ref{E_t_z}) and (\ref{E_l_z}), we 
conclude that field $E_{||}$ should have, in the general case, the 
high-frequency component proportional to $E_{\perp}$. 
Taking this into account, we make use of the next representation 
\begin{eqnarray}
\begin{array}{c}
\sigma=R\,\exp[i\omega(t-z/a)]_{\mathstrut},\\
\displaystyle E_{\perp}=(\Omega_{\perp}/2)^{\mathstrut}_{\mathstrut}
\exp[i\omega(t-z/a)]+\mbox{c.c.},\\ 
\displaystyle E_{||}=(\Omega_{||}-\delta E_{\perp})^{\mathstrut}/2, 
\end{array}
\label{sve}
\end{eqnarray}
where $\omega=\omega_0-\Delta$ is the carrier frequency 
($|\Delta|\ll\omega_0$); $R$, $\Omega_{\perp}$ and $\Omega_{||}$ are slowly 
varying functions of $t$ and $z$ in the standard sense \cite{AE}. 
Substituting these expressions into Eqs. (\ref{s_3_t})--(\ref{E_l_z}) and 
disregarding the nonresonant terms give us following equations 
\begin{eqnarray}
\frac{\partial W}{\partial t}=\frac{i}{2}(\Omega_{\perp}^*R-\Omega_{\perp}R^*),
\label{W_t}
\end{eqnarray}
\begin{eqnarray}
\frac{\partial R}{\partial t}=i(\Delta+\Omega_{||})R+i\Omega_{\perp}W,
\label{R_t}
\end{eqnarray}
\begin{eqnarray}
\frac{\partial\Omega_{\perp}}{\partial z}+
\frac{1}{a}\frac{\partial\Omega_{\perp}}{\partial t}=-i\beta_{\perp}R,
\label{O_t_z}
\end{eqnarray}
\begin{eqnarray}
\frac{\partial\Omega_{||}}{\partial z}+\frac{1}{a}\frac{\partial\Omega_{||}}
{\partial t}=\beta_{||}\frac{\partial W}{\partial t},
\label{O_l_z}
\end{eqnarray}
where 
\[
\beta_{\perp}=\frac{n\omega_0A}{2\hbar\rho a^3},\quad 
\beta_{||}=\frac{nAd^2}{4\hbar\rho a^3}.
\]
Note that we neglect the high-frequency addition to detuning $\Delta$ in the 
right-hand side of Eq.(\ref{R_t}). 
An influence of this term on the dynamics of the pulses, whose duration is 
much greater than the oscillation period, is not significant because its 
average value over the pulse length tends to zero. 
In the case of the RMB equations for anisotropic media \cite{AEGM}, this fact 
was established in Ref.~\onlinecite{SU3}. 

System (\ref{W_t})--(\ref{O_l_z}) we finally obtain differs only by the 
notations from the LSWC equations that have been investigated in details due 
to their importance for theoretical study of the nonlinear dynamics of 
two-component electromagnetic pulses in the anisotropic resonant media 
\cite{S1,SU2}. 
This reveals to us one more correspondence between coherent optical and 
acoustic phenomena, which will be used in the next section for classification 
of the modes of the acoustic pulse propagation through strained paramagnetic 
crystal. 
It should be noted here that the system of LSWC equations, as well as its 
gauge equivalents, arises in various physical problems (see, e.g., references 
in Ref.~\onlinecite{SU2}). 
These numerous applications point to the universal physical character of the 
LSWC system. 

One of the features of Eqs. (\ref{W_t})--(\ref{O_l_z}) is that the field 
variables play in them totally different roles. 
Namely, $\Omega_{\perp}$ causes the quantum transitions, whereas 
$\Omega_{\perp}$ shifts dynamically the transition frequency.  
It is easy to see that there exists relation between the field components: 
\begin{eqnarray}
\Omega_{||}=-\frac{d^2}{4\omega_0}|\Omega_{\perp}|^2+F(t-z/a). 
\label{l_t}
\end{eqnarray}
The last term on the right-hand side generates the phase modulation of $R$ and 
$\Omega_{\perp}$ only and vanishes after appropriate change of the variables. 
Without loss of a generality, we put $F(t-z/a)=0$ in what follows. 

At the end of this section, we find the expressions for transverse and 
longitudinal components of the strain field through the variables of the LSWC 
equations. 
It follows from Eqs. (\ref{new_var}) and (\ref{sve}) that 
\begin{eqnarray}
{\cal E}_{xz}=\frac{\hbar G_{\perp}}{A}\cos2\alpha\Bigl(
\Omega_{\perp}\exp[i\omega(t-z/a)]+\mbox{c.c.}\Bigr)
\nonumber\\ \mbox{}
-\frac{\hbar}{2G_{\perp}}\sin2\alpha\,\,\Omega_{||},
\label{E_xz}
\end{eqnarray}
\begin{eqnarray}
{\cal E}_{zz}=\frac{\hbar G_{||}}{A}\sin2\alpha
\Bigl(\Omega_{\perp}\exp[i\omega(t-z/a)]+\mbox{c.c.}\Bigr)
\nonumber\\ \mbox{}
+\frac{\hbar}{G_{||}}\cos2\alpha\,\,\Omega_{||}.
\label{E_zz}
\end{eqnarray}
These formulas display that both the components of the acoustic pulse have 
high-frequency and zero-frequency components in the general case. 
Also, Eqs. (\ref{l_t})--(\ref{E_zz}) reveal us an asymmetry on the polarity of 
acoustic signal: the signs of the zero harmonics of its transverse and 
longitudinal components are determined by $\alpha$ and the type of the 
external action (tension or compression) on the crystal. 
It is remarkable that the ratio of the amplitudes of the zero harmonic of 
${\cal E}_{zz}$ and ${\cal E}_{xz}$\,:
\[
\frac{G_{\perp}}{G_{||}}\cot2\alpha,
\]
is equal to inverse ratio of the first harmonic case. 
This fact can be used in measuring the constants of the spin-phonon 
interaction in the paramagnetic crystals. 

\section{THE MODES OF ACOUSTIC TRANSPARENCY}

The solution of Eqs. (\ref{W_t})--(\ref{O_l_z}) that describes the propagation 
of the transverse-longitudinal acoustic pulse is written as given (see 
Ref.~\onlinecite{SU2}): 
\begin{eqnarray}
\Omega_{\perp}=\sqrt{M}\exp(i\Phi),
\label{O_t}
\end{eqnarray}
\begin{eqnarray}
\Omega_{||}=-\frac{d^2}{4\omega_0}\,M,
\label{O_l}
\end{eqnarray}
\begin{eqnarray}
W=\left(1-\frac{\tau_p^2}{2(1+\alpha^2)}\,M\right)W_0.
\label{W}
\end{eqnarray}
Here  
\begin{eqnarray}
M=\frac{8g}{\tau_p^2\left(g-\alpha+\sqrt{1+(g-\alpha)^2}
\cosh2\zeta\right)^{\mathstrut}},
\label{M}
\end{eqnarray}
\begin{eqnarray}
\Phi=\frac{\beta_{\perp}\alpha\tau_p}{1+\alpha^2}\,W_0z-\mbox{arctan}
\frac{\tanh\zeta}{s}+{\rm const},
\label{Phi}
\end{eqnarray}
\[
\alpha=\Delta\tau_p,\quad W_0=\frac{1-3W_2}{4},
\]
\[
g=\frac{2\omega_0\tau_p}{d^2},\quad
\zeta=\tau_p^{-1}\left(t-\frac{z}{v_g}\right),
\]
\[
s=g-\alpha+\sqrt{1+(g-\alpha)^2},
\]
\begin{eqnarray}
v_g=a\left(1-\frac{a\beta_{\perp}\tau_p^2}{1+\alpha^2}W_0\right)^{-1}.
\label{v_g}
\end{eqnarray}
The free parameters of the pulse presented are $\Delta$ and $\tau_p$.  
For the sake of convenience we suppose hereinafter that $\tau_p>0$. 

As it follows from Eqs. (\ref{O_t}) and (\ref{Phi}), component 
$\Omega_{\perp}$ has the phase modulation, which leads to a local nonlinear 
chirping of the carrier frequency, 
$\omega\to\omega_{loc}=\omega+\delta\omega_{non}$, where 
\begin{eqnarray}
\delta\omega_{non}\equiv\frac{\partial\Phi}{\partial t}=\frac{\Omega_{||}}{4}.
\label{d_o_n}
\end{eqnarray}
Taking into account the dynamic shift of the transition frequency 
$\omega_0\to\omega_0^{ef}=\omega_0+\Omega_{||}$ (see Eq. (\ref{R_t})), we come 
to the following expression for effective detuning $\Delta_{ef}$ of component 
$\Omega_{\perp}$ from the resonance: 
\begin{eqnarray}
\Delta_{ef}\equiv\omega_0^{ef}-\omega_{loc}=\Delta+\frac34\Omega_{||}. 
\label{d_ef}
\end{eqnarray}

Defining the pulse length $T_p$ as the double deviation from the zero point of 
$t-z/v_g$, at which $|\Omega_{\perp}|$ is half its maximum value, we obtain 
from formula (\ref{O_t}): 
\begin{eqnarray}
T_p=\tau_p\,{\rm arccosh}\left(4+3\frac{g-\alpha}{\sqrt{1+(g-\alpha)^2}}
\right).
\label{T_p}
\end{eqnarray}
In the SVE approximation, both pulse length and nonlinear shift of the carrier 
frequency must obviously satisfy the conditions $\omega_0T_p\gg1$ and 
$|\delta\omega_{non}|\ll\omega_0$. 
It can easily be shown that these inequalities are valid if 
$\omega_0\tau_p\gg1$ and $\omega_0\tau_p\gg\alpha-g$. 
The last condition is necessary in the case $\alpha-g\gg1$, evidently, and can 
be fulfilled only when $d^2\gg1$. 

Considering the limit $\tau_p\to\infty$ in formulas (\ref{O_t})--(\ref{v_g}) 
gives us rationally decreasing pulse solution: 
\begin{eqnarray}
\Omega_{\perp}=\frac{8i\omega_0\kappa}{d^2(1+i\kappa^2\xi)}
\exp(i\beta_{\perp}W_0z/\Delta),
\label{O_t_r}
\end{eqnarray}
\begin{eqnarray}
W=\left(1-\frac{8\kappa^2}{(1+\kappa^2)^2(1+\kappa^4\xi^2)}\right)W_0,
\label{W_r}
\end{eqnarray}
where
\[
\xi=\frac{4\omega_0}{d^2}\left(t-\frac{z}{v_r}\right),\quad
\kappa=\sqrt{\frac{d^2}{2\omega_0}\Delta-1},
\]
\[
v_r=a\left(1-\frac{a\beta_{\perp}}{\Delta^2}W_0\right)^{-1}.
\]
Parameter $\kappa$ in Eqs. (\ref{O_t_r}) and (\ref{W_r}) is supposed to be 
real.
This imposes a constraint on detuning $\Delta$ of the rationally decreasing
pulses: $\Delta>\Delta_r$, where $\Delta_r=2\omega_0/d^2$.

Now, we use the expressions presented above to identify the modes of the 
propagation of transverse-longitudinal acoustic pulses in deformed 
paramagnetic crystal. 
These modes resemble the ones studied in Ref.~\onlinecite{SU2} for the case of 
two-component electromagnetic pulses. 
In what follows, it is assumed for the sake of concreteness that the 
paramagnetic impurities are in thermodynamic equilibrium prior to the pulse 
passage ($-1/2<W_0<0$). 

Let us begin with the modes, in which the pulse excites the paramagnetic 
impurities strongly. 
Since $W^2+|R|^2$ is independent of $t$, the strongest degree of excitation 
happens in the case, when the value of variable $W$ in the pulse center 
differs by the sign from its value at absence of the pulse. 
By virtue of Eqs. (\ref{W}) and (\ref{M}), this condition can be formulated as 
\[
\tau_p^2|\Omega_{\perp}(\zeta=0)|^2=4(1+\alpha^2). 
\]
Performing some simple algebra, we put it into the next form 
\[
(\alpha^2-2g\alpha+1)^2=0, 
\] 
which yields
\begin{eqnarray}
g=\frac12\left(\alpha+\frac{1}{\alpha}\right). 
\label{se_g}
\end{eqnarray}
This condition can take place if $\alpha>0$ only. 
It can also be rewritten as a relation between the pulse detuning $\Delta$ and 
parameter $\tau_p$\,:
\begin{eqnarray}
\tau_p^{-1}=\sqrt{\frac{4\omega_0}{d^2}\Delta-\Delta^2}
\label{se}
\end{eqnarray}
Setting the right-hand side to zero, we find the interval of admissible values 
of the detuning: $0<\Delta<\Delta_m=2\Delta_r$.  
The curve of the strong excitation (\ref{se}) is plotted in Fig.~3.
\begin{figure}[ht]
\centering
\includegraphics[angle=0,width=3.0in]{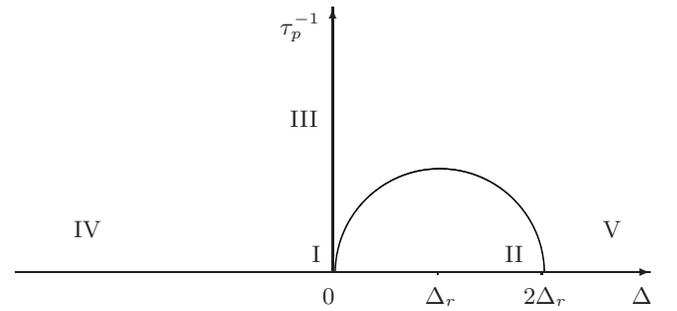}
\begin{picture}(0,0)
\put(-232,61.9){\vector(1,0){240}}
\put(-112.1,62){\vector(0,1){100}}
\put(-120,65.5){\small I}
\put(-47,65.5){\small II}
\put(-128.2,117){\small III}
\put(-210,75){\small IV}
\put(-10,75){\small V}
\put(1,50){$\Delta$}
\put(-116,50){0}
\put(-40,50){$2\Delta_r$}
\put(-33.0,61.9){\line(0,-1){1}}
\put(-77.2,50){$\Delta_r$}
\put(-72.5,61.9){\line(0,-1){1}}
\put(-132,152){$\tau_p^{-1}$}
\end{picture}
\vskip-1.4cm
\caption{Curve of the strong excitation. Domains of the pulse parameters 
corresponding to different modes of acoustic transparency: (I) ASIT; (II) 
ASIST; (III) AEOT; (IV) ANNT; (V) APNT.} 
\end{figure}

It is seen from Eq. (\ref{W_r}) that the strongest excitation in the case of 
rationally decreasing pulse takes place if $\kappa=1$. 
Obviously, this agrees with (\ref{se}). 

Substitution of relation (\ref{se_g}) into Eqs. (\ref{O_t})--(\ref{v_g}) 
yields 
\begin{eqnarray}
\begin{array}{c}
\displaystyle|\Omega_{\perp}|=\frac{2\sqrt{1+\alpha^2}}
{\tau_p\sqrt{1+(1+\alpha^2)\sinh^2\zeta_{\mathstrut}}},\\
\displaystyle\Omega_{||}=-\frac{4\Delta}
{1+(1+\alpha^2)\sinh^2\zeta}^{\mathstrut}_{\mathstrut},\\
\displaystyle W=\left(1-\frac{2}{1+(1+\alpha^2)\sinh^2\zeta}
\right)^{\mathstrut}_{\mathstrut}W_0,\\
\displaystyle 
v_g=a{\left(1-\frac{a\beta_{\perp}\tau_p^2}{1+\alpha^2}W_0
\right)^{-1}}^{\mathstrut}.
\end{array}
\label{se_f}
\end{eqnarray}
The corresponding pulse length is 
\begin{eqnarray}
T_p=2\tau_p\,{\rm arcsinh}\sqrt{\frac{3}{1+\alpha^2}}. 
\label{t_s}
\end{eqnarray}
Accordingly to Eqs. (\ref{d_ef}) and (\ref{se_f}), the effective detuning of 
$\Omega_{\perp}$ from the resonance equals $\Delta$ at the edges of the pulse 
and $-2\Delta$ at its center. 
Thus, we can say that this component is resonant with quantum transitions on
average over the pulse length. 
As a result, the largest possible change in the population of the spin 
sublevels is achieved. 

If the detuning is very small ($\alpha\ll1$, $g\gg1$), then Eqs. (\ref{se_f}) 
can be represented as 
\begin{eqnarray}
\begin{array}{c}
\displaystyle|\Omega_{\perp}|=\frac{2}{{\tau_p}_{\mathstrut}}\,
\mbox{\rm sech}\zeta,\quad\Omega_{||}=-4\Delta\,\mbox{\rm sech}^2\zeta,\\
\displaystyle W=\left(1-2\,\mbox{\rm sech}^2\zeta
\right)^{\mathstrut}_{\mathstrut}W_0,\\
\displaystyle v_g=a{\left(1-a\beta_{\perp}\tau_p^2W_0
\right)^{-1}}^{\mathstrut}.
\end{array}
\label{se_f_a}
\end{eqnarray}
One can see that $|\Omega_{||}|\ll|\Omega_{\perp}|$, and the phase-modulation 
depth for component $\Omega_{\perp}$ is much smaller than its input spectral 
width: $|\delta\omega_{non}|\ll1/\tau_p$. 
In the exact resonance case ($\alpha=0$, $g\to\infty$), these relations give 
us well-known expressions of the ASIT theory \cite{Sh,D,SSSh}. 

Formulas (\ref{se_f}) and (\ref{d_o_n}) show that an increase of detuning (and 
therefore $\alpha$) leads to larger amplitudes of the pulse components and to 
a deeper chirping of $\Omega_{\perp}$ toward lower frequencies. 
Since the pulse length decreases, the profiles of both components become 
sharper. 

Also, it is necessary to note that the group velocity approaches linear 
velocity $a$ of acoustic wave as detuning increases. 
Nevertheless, the strong excitation of the paramagnetic impurities takes 
place: the largest possible change in the level population is reached at 
the center of the pulse. 
This is evidently explained by the fact that a growth in the amplitude of 
pulse increases its power. 
The ensuing higher rate of excitation/de-excitation processes leads to a 
higher soliton propagation velocity.

It is clear from (\ref{se_g}) that each particular value of $g$ corresponds to 
two distinct values of $\alpha$.  
We associate the domain of relatively small detuning ($\alpha<1$) with the 
ASIT mode, since it is implemented in the case $\alpha=0$. 
When $\alpha>1$, we say that the pulse solutions given by (\ref{se_f}) 
propagate in the mode of acoustic self-induced supertransparency (ASIST), 
thus emphasizing the fact that the group velocity decrease is lower as 
compared to the ASIT mode while excitation is equally strong. 
Since $|\Delta_{ef}|\ll\omega_0$ in the SVE approximation, the ASIST mode must 
be most strongly manifested if $d^2\gg1$. 
The domains of the values of the pulse parameters, which correspond to these 
modes, are schematized in Fig.~3. 

When $g\gg1$ and $\alpha\gg1$, the amplitudes of components $\Omega_{\perp}$ 
and $\Omega_{||}$ of the ASIST pulse and the phase-modulation depth for 
$\Omega_{\perp}$ reach their limits $2\Delta_m$, $4\Delta_m$ and $\Delta_m$, 
respectively. 
In this case, (\ref{t_s}) gives us the following estimate for the pulse length:
\[
T_p\approx\frac{2\sqrt3\tau_p}{\alpha}\approx\frac{d^2}{\omega_0}.
\]
This time scale corresponds to the time scale of phase-modulation 
localization. 
A deeper phase modulation combined with a decrease in the corresponding 
localization time scale and a shorter pulse length can be interpreted as an 
effect of the spectral supercontinuum generation. 
Indeed, the Fourier transform of (\ref{O_t}) and (\ref{O_l}) defined as 
\[
F_{\perp,||}(\nu)=\int\limits_{-\infty}^{\infty}\mbox{e}^{\displaystyle i\nu t}
\Omega_{\perp,||}\,dt
\]
yields 
\[
|F_{\perp}(\nu)|=2\pi\frac{\sqrt{g}}{\sqrt[4\,\,]{1+(g-\alpha)^2}}\,
\frac{\mbox{e}^{\displaystyle\theta\tau_p\nu/2}}{\cosh(\pi\tau_p\nu/2)}, 
\]
\[
|F_{||}(\nu)|=4\pi\frac{\sinh(\theta\tau_p\nu/2)}{\sinh(\pi\tau_p\nu/2)}, 
\]
where $\theta=\mbox{arccot}(g-\alpha)$ ($0<\theta<\pi$). 
The absolute values of the Fourier transforms of the components $\Omega_{||}$ 
and $\Omega_{\perp}$ reach their maximum values at $\nu=0$ and $\nu=\nu_0$, 
respectively, where
\[
\nu_0=\frac{2}{\pi\tau_p}\ln\frac{\pi+\theta}{\pi-\theta}. 
\]
If $\alpha-g\gg1$ or, equivalently, $\theta\to\pi$, then the spectral width of 
the pulse is $\delta\omega\sim1/(\pi-\theta)\tau_p$ and the maximum of the 
spectral energy distribution of component $\Omega_{\perp}$ is reached at 
frequency $\omega-\nu_0$, which is much lower than the carrier frequency 
$\omega$. 
In this case, despite a large linear detuning from resonance 
($\tau_p\Delta\gg1$), a substantial nonlinear spectral broadening 
($\tau_p\delta\omega\gg1$) and dynamical shift in the transition frequency 
$\omega_0\to\omega_0^{ef}$ lead to a generation of the resonant Fourier 
components (photons), which stimulate the quantum transitions. 

Let us now consider the modes of the acoustic pulse propagation in the case 
$g\ll1$. 
Although the paramagnetic impurities are excited weakly under this condition, 
there exist remarkable features in their interaction with the acoustic pulses. 
Since $\omega_0\tau_p\gg1$ in the SVE approximation, the value of $|d|$ must 
be sufficiently large to ensure that $\omega_0\tau_p/d^2\ll1$. 
This can be fulfilled for some values of $\alpha$ if $|G_{\perp}|\gg|G_{||}|$ 
or vice versa. 

First, we assume that the detuning from resonance is large ($|\alpha|\gg1$). 
If $\alpha<0$ (or $\omega>\omega_0$), then Eqs. (\ref{O_t})--(\ref{v_g}) give 
\begin{eqnarray}
\begin{array}{c}
\displaystyle|\Omega_{\perp}|=\frac{2}{\tau_p}
{\sqrt{\frac{g}{|\alpha|}}}_{\mathstrut}\,\mbox{\rm sech}\,\zeta,\\
\displaystyle\Omega_{||}=-{\frac{2}{|\alpha|\tau_p}}_{\mathstrut}^{\mathstrut}
\,\mbox{\rm sech}^2\zeta,\\
\displaystyle W=\left(1-\frac{2g}{|\alpha|^3}\,\mbox{\rm sech}^2\zeta
\right)_{\mathstrut}^{\mathstrut}W_0,\\
v_g=v_r^{\mathstrut}.
\end{array}
\label{nnt}
\end{eqnarray}
Comparing these expressions with (\ref{se_f_a}), we see that the amplitude of 
$\Omega_{\perp}$ is much smaller than in the ASIT mode, whereas the respective 
$\Omega_{||}$'s are comparable. 
We also note that
\[
\frac{|\Omega_{\perp}(\zeta=0)|}{|\Omega_{||}(\zeta=0)|}=\sqrt{g|\alpha|},
\]
i.e., the amplitude ratio can have an arbitrary value. 
The paramagnetic impurities remain almost unexcited as the soliton described 
by (\ref{nnt}) propagates through the crystal, and the soliton velocity 
decreases only very slightly. 
The phase-modulation depth is also small ($|\delta\omega_{non}|\ll1/\tau_p$), 
and the effective detuning $\Delta_{ef}$ only increases as the component 
$\Omega_{||}$ is generated. 
This leads to an even weaker excitation of the impurities as compared to that 
induced by the input pulse. 

Now, let us suppose that $g\ll1$ and $\alpha\gg1$. 
In this case Eqs. (\ref{O_t})--(\ref{v_g}) imply 
\begin{eqnarray}
\begin{array}{c}
\displaystyle|\Omega_{\perp}|={\frac{4\sqrt{g\alpha}}
{\tau_p\sqrt{1+4\alpha^2\sinh^2\zeta}}}_{\mathstrut},\\
\displaystyle\Omega_{||}=-{\frac{8\Delta}{1+4\alpha^2\sinh^2\zeta}
}_{\mathstrut}^{\mathstrut},\\
\displaystyle W=\left(1-\frac{8g}{\alpha(1+4\alpha^2\sinh^2\zeta)}
\right)_{\mathstrut}^{\mathstrut}W_0,\\
v_g=v_r^{\mathstrut}.
\end{array}
\label{pnt}
\end{eqnarray}

Note that expressions (\ref{se_f}) and (\ref{pnt}) are somewhat similar: in 
both cases, the solitons are sharply peaked, and their propagation velocities 
are nearly equal to $a$. 
However, these modes are essentially different in terms of behavior of the 
paramagnetic impurities. 
Whereas they are strongly excited as the soliton described by (\ref{se_f}) 
propagates through crystal, no significant excitation is caused in the case of 
(\ref{pnt}). 
Indeed, since the effective detuning $\Delta_{ef}$ of the pulse described by 
Eqs. (\ref{pnt}) is $-5\Delta$ at its center, it is not resonant with the 
paramagnetic impurities on average over the pulse length. 
However, according to (\ref{nnt}) and (\ref{pnt}), the excitation of the 
paramagnetic impurities at $\omega<\omega_0$ being relatively weak, is still 
stronger than that at $\omega>\omega_0$. 
The reason is that the effective detuning decreases toward the pulse center 
when $\omega<\omega_0$, owing to the component $\Omega_{||}$, and increases 
when $\omega>\omega_0$. 
Thus, a comparison of (\ref{nnt}) with (\ref{pnt}) demonstrates obvious 
asymmetry with respect to detuning of $\Omega_{\perp}$. 
Since $\alpha<0$ for the pulses described by (\ref{nnt}) and $\alpha>0$ for 
ones described by (\ref{pnt}), we refer to the corresponding modes as acoustic 
negative and positive nonresonant transparency (ANNT and APNT), respectively. 

If $g\ll1$ and $|\alpha|\ll1$ (detuning is small), then 
Eqs. (\ref{O_t})--(\ref{v_g}) lead to expressions identical to those 
found in Ref.~\onlinecite{S1}: 
\begin{eqnarray}
\begin{array}{c}
\displaystyle|\Omega_{\perp}|={\frac{2}{\tau_p}}_{\mathstrut}
\sqrt{2g}\,\mbox{\rm sech}^{1/2}2\zeta,\\
\displaystyle\Omega_{||}=-{\frac{4}{\tau_p}}_{\mathstrut}^{\mathstrut}
\,\mbox{\rm sech}\,2\zeta,\\
W=(1-4g\,\mbox{\rm sech}\,2\zeta)^{\mathstrut}W_0.
\end{array}
\label{aet}
\end{eqnarray}
The expression for group velocity coincides with one given in Eqs. 
(\ref{se_f_a}) corresponding to the ASIT mode. 

Here, we have $|\Omega_{||}/\Omega_{\perp}|^2\gg1$ in the center of the pulse. 
The paramagnetic impurities are not excited, since the effective detuning is 
large, $3/\tau_p$. 
However, the propagation velocity decreases as in the case of strong 
excitation at $\alpha\sim1$.
Such a deceleration of the pulse is explained by the dispersion properties of 
medium within the higher-frequency component bandwidth. 
According to \cite{S1}, where this effect was studied in details for the 
optical solitons, we call the mode considered an acoustic extraordinary 
transparency (AEOT). 
The roles of the ordinary and extraordinary pulse components are played here 
by $\Omega_{\perp}$ and $\Omega_{||}$, respectively. 
The existence of this mode for transverse-longitudinal acoustic pulses has 
been revealed also in the case, when the splitting of the spin sublevels is 
produced by the external magnetic field \cite{VS,GS}. 
The domains of existence of the modes with $g\ll1$ are presented in Fig.~3. 

It should be noted that expressions (\ref{O_t})--(\ref{v_g}) do not change 
their form in a presence of detuning between $a_{||}$ and $a_{\perp}$. 
The influence of this detuning is inessential if condition 
\[
\varepsilon\equiv\left(\frac{1}{a_{\perp}}-\frac{1}{a_{||}}\right)
\left(\frac{1}{v_g}-\frac{1}{a}\right)^{-1}\ll1
\]
is valid. 
Substituting the definitions for $\beta_{\perp}$ and $A$ into (\ref{v_g}), we 
obtain  
\[
\varepsilon\sim\frac{1}{a\beta_{\perp}\tau_p^2}\left(\frac{1}{a_{\perp}}-
\frac{1}{a_{||}}\right)\sim\frac{\hbar\rho a_{||}^2}{nG_{||}^2\omega_0\tau_p^2}
\left(\frac{a_{||}}{a_{\perp}}-1\right). 
\]
Taking for $\rm Fe^{2+}$:$\rm MgO$ \cite{Sh,TR,S2,P,S4} 
$n\sim10^{17}\mbox{\,cm}^{-3}$, $\omega_0\sim10^{10}\mbox{\,s}^{-1}$, 
$G_{||}\sim10^{-13}\mbox{\,erg}$, $\rho\approx2\mbox{\,g}/\mbox{cm}^3$, 
$a_{||}\sim5\cdot10^5\mbox{\,cm/s}$, $a_{||}/a_{\perp}\sim1.5$ and 
$\tau_p\sim10^{-8}\mbox{\,s}$, we find $\varepsilon\sim0.1$. 
This estimation shows that we can neglect detuning between the linear 
velocities of transverse and longitudinal acoustic waves for typical values of 
the ASIT parameters. 

\section{CONCLUSION} 

In this paper, we have investigated the soliton modes of the acoustic 
transparency in a strained cubic crystal containing the resonant paramagnetic 
impurities with effective spin $S=1$. 
It is supposed that the linear velocities of transverse and longitudinal sound 
are close, and the pulses propagate through the crystal under arbitrary angle 
with respect to the direction of external deformation parallel to the 
fourth-order symmetry axis. 

We have allocated five modes of acoustic transparency, which differ by the 
propagation velocity of the transverse-longitudinal pulses and degree of 
excitation of the paramagnetic impurities (see Table~I). 
\begin{table}
\caption{Characteristics of the modes of acoustic transparency}
\begin{tabular}{p{1.0cm}|p{1.0cm}|p{0.9cm}|p{0.9cm}|p{1.3cm}|p{1.3cm}|p{1.1cm}}
\hline Mode \hfill\mbox{}&
\hfill $\tilde\Delta_{ef}$\footnote{Note: $\tilde\Delta_{ef}$ is the 
effective detuning of the high-frequency components at the center of acoustic 
pulse; $\tilde\Delta_{ef}$ is compared with $\Delta$ in the AEOT mode on 
absolute value.}
\hfill\mbox{}&
\hfill $|\alpha|$\hfill\mbox{}&
\hfill $g$\hfill\mbox{}&
\hfill $v_g$\hfill\mbox{}& 
\hfill $|\Omega_{\perp}/\Omega_{||}|$\hfill\mbox{}&
\hfill\parbox{1.05cm}{\hfill $\mbox{Excita-}^{\mathstrut}$\hfill\mbox{}\\ 
\hfill $\mbox{tion}_{\mathstrut}$\hfill\mbox{}}\hfill\mbox{}\\ 
\hline
\vskip-0.2cm ASIT\hfill\vphantom{$\Omega_M v_{SIT}$}&
\vskip-0.2cm\hfill $-2\Delta$\hfill\vphantom{$\Omega_M v_{SIT}$}&
\vskip-0.2cm\hfill $\ll1$\hfill\vphantom{$\Omega_M v_{SIT}$}&
\vskip-0.2cm\hfill $\gg1$\hfill\vphantom{$\Omega_M v_{SIT}$}&
\vskip-0.2cm\hfill $v_{ASIT}$\hfill\vphantom{$\Omega_M v_{SIT}$}&
\vskip-0.2cm\hfill $\gg1$\hfill\vphantom{$\Omega_M v_{SIT}$}&
\vskip-0.2cm strong\hfill\vphantom{$\Omega_M v_{SIT}$}\\ 
\hline
\vskip-0.2cm ASIST\hfill\vphantom{$\Omega_M v_{SIT}$}&
\vskip-0.2cm\hfill $-2\Delta$\hfill\vphantom{$\Omega_M v_{SIT}$}&
\vskip-0.2cm\hfill $\gg1$\hfill\vphantom{$\Omega_M v_{SIT}$}&
\vskip-0.2cm\hfill $\gg1$\hfill\vphantom{$\Omega_M v_{SIT}$}&
\vskip-0.2cm\hfill $<a$\hfill\vphantom{$\Omega_M v_{SIT}$}&
\vskip-0.2cm\hfill $\sim1$\hfill\vphantom{$\Omega_M v_{SIT}$}&
\vskip-0.2cm strong\hfill\vphantom{$\Omega_M v_{SIT}$}\\ 
\hline
\vskip-0.2cm ANNT\hfill\vphantom{$\Omega_M v_{SIT}$}&
\vskip-0.2cm\hfill $\approx\Delta$\hfill\vphantom{$\Omega_M v_{SIT}$}&
\vskip-0.2cm\hfill $\gg1$\hfill\vphantom{$\Omega_M v_{SIT}$}&
\vskip-0.2cm\hfill $\ll1$\hfill\vphantom{$\Omega_M v_{SIT}$}&
\vskip-0.2cm\hfill $a$\hfill\vphantom{$\Omega_M v_{SIT}$}&
\vskip-0.2cm\hfill arbitrary\hfill\vphantom{$\Omega_M v_{SIT}$}&
\vskip-0.2cm weak\hfill\vphantom{$\Omega_M v_{SIT}$}\\
\hline
\vskip-0.2cm APNT\hfill\vphantom{$\Omega_M v_{SIT}$}&
\vskip-0.2cm\hfill $-5\Delta$\hfill\vphantom{$\Omega_M v_{SIT}$}&
\vskip-0.2cm\hfill $\gg1$\hfill\vphantom{$\Omega_M v_{SIT}$}&
\vskip-0.2cm\hfill $\ll1$\hfill\vphantom{$\Omega_M v_{SIT}$}&
\vskip-0.2cm\hfill $a$\hfill\vphantom{$\Omega_M v_{SIT}$}&
\vskip-0.2cm\hfill $\ll1$\hfill\vphantom{$\Omega_M v_{SIT}$}&
\vskip-0.2cm weak\hfill\vphantom{$\Omega_M v_{SIT}$}\\
\hline
\vskip-0.2cm AEOT\hfill\vphantom{$\Omega_M v_{SIT}$}&
\vskip-0.2cm\hfill $\gg|\Delta|$\hfill\vphantom{$\Omega_M v_{SIT}$}&
\vskip-0.2cm\hfill $\ll1$\hfill\vphantom{$\Omega_M v_{SIT}$}&
\vskip-0.2cm\hfill $\ll1$\hfill\vphantom{$\Omega_M v_{SIT}$}&
\vskip-0.2cm\hfill $>v_{ASIT}$\hfill\vphantom{$\Omega_M v_{SIT}$}&
\vskip-0.2cm\hfill $\ll1$\hfill\vphantom{$\Omega_M v_{SIT}$}&
\vskip-0.2cm weak\hfill\vphantom{$\Omega_M v_{SIT}$}\\
\hline
\end{tabular}
\end{table}
The acoustic self-induced transparency is characterized by strong excitation 
and substantial deceleration in the pulse propagation velocity relative to 
linear velocities. 
Self-induced supertransparency differs from ASIT in that the decrease of 
velocity is small, but the paramagnetic impurities are strongly excited as 
well. 
The solitons of ASIST have larger amplitudes and smaller lengths as compared 
to ASIT solitons, and their high-frequency spectral components are strongly 
modulated. 
The carrier frequency of the transverse and longitudinal components in this 
mode is lower than the resonant frequency. 
The modes, in which the trapping of the populations of the spin sublevels 
takes place, are also identified. 
The pulses propagating in the acoustic extraordinary transparency mode are 
characterized by small detuning of the high-frequency components and dominant 
role of the zero-frequency ones. 
Their group velocity substantially changes and may become comparable to that
of pulses in ASIT and ASIST modes. 
In the acoustic positive and negative nonresonant transparency modes, the 
pulse velocity changes insignificantly, and the absolute value of detuning is 
large. 
The most substantial difference between these modes concerns the behavior of 
the effective detuning of the high-frequency components. 
In the ANNT mode, it remains virtually constant. 
If a pulse propagates in the APNT mode, then the effective detuning changes 
sign due to an influence of the zero-frequency component. 
Since the local frequency passes through a resonance, a slightly stronger 
excitation occurs in this case, and the pulses are sharply peaked, as in the 
ASIST mode. 

In this study, we ignored inhomogeneous broadening of the spin sublevels. 
An investigation allowing for this effect may throw light on the pulse area 
theorem, providing a basis for analysis of the stability of the 
transverse-longitudinal acoustic pulses. 
Also, it would be interesting to identify the distinctive features of the 
dynamics of picosecond transverse-longitudinal acoustic pulses in a system of 
effective spins $S=1$. 
Unlike the present consideration, the SVE approximation is inapplicable to 
this case. 

\section*{ACKNOWLEDGMENT}

This study was supported by the Russian Foundation for Basic Research (Grant 
No. 05-02-16422).

\vfill
\eject
\end{document}